\documentclass[fleqn,usenatbib]{mnras}
\usepackage{newtxtext,newtxmath}
\usepackage[T1]{fontenc}
\DeclareRobustCommand{\VAN}[3]{#2}
\let\VANthebibliography\thebibliography
\def\thebibliography{\DeclareRobustCommand{\VAN}[3]{##3}\VANthebibliography}

\usepackage{graphicx}	
\usepackage{amsmath}	

\usepackage{multirow}
\usepackage{braket}
\usepackage{comment}
\usepackage{bm}
\usepackage{graphicx}
\usepackage{soul}

\usepackage[dvipsnames]{xcolor}
\usepackage[capitalise]{cleveref}
\crefformat{equation}{Eq. (#2#1#3)}
\crefrangeformat{equation}{Eqs. #3#1#4-#5#2#6}
\crefrangeformat{figure}{Figs. #3#1#4-#5#2#6}
\Crefformat{equation}{Equation (#2#1#3)}
\Crefname{equation}{Equation}{Equations}

\title[Point-cloud diffusion galaxy mocks]
{From Dark Matter to Galaxies: Halo-Free Mock Generation via Conditional Point-Cloud Diffusion}

\author[K. Moriwaki et al.]{
Kana Moriwaki,$^{1,2}$\thanks{E-mail: kana.moriwaki@phys.s.u-tokyo.ac.jp}, 
Ken Osato,$^{3,4,2,6}$
Naoki Yoshida$^{2,5,6}$
\\
$^{1}$Research Center for the Early Universe, Graduate School of Science, The University of Tokyo, 7-3-1 Hongo, Bunkyo, Tokyo 113-0033, Japan\\
$^{2}$RIKEN Center for Advanced Intelligence Project, 1-4-1 Nihonbashi, Chuo, Tokyo 103-0027, Japan\\
$^{3}$Center for Frontier Science, Chiba University, 1-33 Yayoi-cho, Inage-ku, Chiba 263-8522, Japan\\
$^{4}$Department of Physics, Graduate School of Science, Chiba University, 1-33 Yayoi-cho, Inage-ku, Chiba 263-8522, Japan\\
$^{5}$Department of Physics, Graduate School of Science, The University of Tokyo, 7-3-1 Hongo, Bunkyo, Tokyo 133-0033, Japan\\
$^{6}$Kavli Institute for the Physics and Mathematics of the Universe, The University of Tokyo Institutes for Advanced Study,\\
5-1-5 Kashiwanoha, Kashiwa-shi, Chiba, 277-8583, Japan\\
}

\date{Accepted XXX. Received YYY; in original form ZZZ}
\pubyear{2026}

\begin{document}
\label{firstpage}
\pagerange{\pageref{firstpage}--\pageref{lastpage}}
\maketitle

\begin{abstract}
We present a diffusion-based generative model for constructing realistic galaxy catalogues from dark matter density fields. 
The model takes a three-dimensional dark matter density field as input and generates galaxies directly as a point cloud with positions and physical properties, including star formation rate (SFR), without identifying dark matter haloes or subhaloes. 
We train the model on galaxy catalogues from the IllustrisTNG hydrodynamical simulation. The generated catalogues reproduce the spatial correspondence between galaxies and the underlying dark matter field, preferentially populating dense regions and filamentary structures. 
They also accurately reproduce the one- and two-dimensional distributions of SFR and stellar mass, the galaxy auto-power spectrum, and the galaxy--dark matter cross-power spectrum. 
The model generates galaxies associated with structures below the nominal resolution of the input density field by marginalising over unresolved small-scale structure rather than relying on a resolved halo catalogue. 
With an optimised diffusion sampling schedule, it generates a catalogue with ${\rm SFR} > 1~{\rm M_\odot/yr}$ over a $(151.3~{\rm Mpc})^3$ volume in approximately 10 seconds on a single GPU. 
Our model therefore provides a practical engine for producing large mock ensembles for upcoming galaxy redshift surveys and line-intensity mapping experiments, and offers a path toward simulation-based inference that bypasses halo finding and directly connects field-level dark matter statistics to observable galaxy populations.
\end{abstract}



\section{Introduction}

Galaxies trace the underlying matter distribution in the Universe and the formation process is affected by
local and large-scale dark matter distributions.
Ongoing and upcoming galaxy redshift surveys \citep[e.g.][]{Takada14, DESI16, Euclid25, LSST09, Roman15} and line-intensity mapping observations \citep[e.g.][]{Crites14, Aguirre18, Concerto20, EXCLAIM20, CCAT-Prime23} will measure the galaxy distribution over increasingly large cosmological volumes. These observations provide crucial information on cosmology as well as galaxy formation and evolution. Generating realistic mock galaxy catalogues is an essential path toward simulation-based inference and
toward statistical analysis of data from the large observational programmes.

Hydrodynamics simulation follows the gas dynamics and models star formation and feedback processes in a physically motivated manner, and thus is currently one of the most reliable methods for generating realistic mocks \citep{Osato2023}. However, when simulating large volumes at high resolution, such direct simulations become computationally expensive, making it impractical to produce a large number of mocks required for statistical analyses. 
An alternative approach is to adopt a heuristic relation between dark matter (DM) halos and baryonic components to {\it paint} the baryonic properties onto the outputs of relatively cheap, DM-only simulations. 

Fast, yet detailed baryon painting can be performed using modern machine learning techniques. 
Several studies have already explored machine learning techniques to {\it paint} galaxies onto DM haloes \citep[e.g.][]{Kamdar16,Jo19,Jespersen22}. 
Deterministic regression models are often employed in these studies, and more recent studies have begun to explore probabilistic generative models as an alternative \citep[e.g.][]{Lovell23, Nguyen24, Moriwaki26}, making it possible to deal with stochastisity in galaxy formation processes and to generate diverse galaxy populations.

Models based solely on the halo-galaxy connection suffer from two limitations. First, it cannot fully capture the complex environmental effects. Although summary statistics such as the local mean density around each halo can be used as input \citep[e.g.][]{Jo19}, they provide only abstracted information. 
Second, this halo-based approach is unable to comprehensively model correlations between the generated galaxies. Some existing methods can account for the correlations among galaxies within individual haloes \citep{Nguyen24, Moriwaki26}, but capturing the correlations across different haloes remains beyond their ability.

An alternative approach for incorporating environmental effects is to generate galaxy ``maps'' from DM density fields \citep{Zhang19, Sether24}, for which convolutional neural networks (CNNs) are commonly employed to process uniformly pixelised or voxelised cubic volumes.
A further advantage of this approach is that even approximate DM density fields obtained from low-cost simulations, such as particle-mesh (PM) simulations \citep[e.g.,][]{Li2022}, can be used as a conditional input.
In such a map-to-map approach, however, there remains a limitation in modelling individual galaxies. 
To identify galaxies as distinct objects on a grid, the spatial resolution must be high enough to separate close pairs, so that each pixel contains at most one galaxy. This is inefficient because the required resolution is set by clustered regions, whereas most pixels remain empty, thereby wasting computational resources.

In the present paper, we study a diffusion model that generates galaxies as a point cloud conditioned on the DM density map. 
A closely related approach has recently been proposed by \citet{Pandey25}, who demonstrated that galaxy point clouds and their physical properties can be generated from three-dimensional DM fields using a transformer-based model. While their study provides an important proof of concept in relatively small periodic simulation volumes, our focus is on scalable generation over much larger volumes. We therefore generate galaxies patch by patch and stitch the generated patches together, rather than modelling a single simulation volume at once. 
This strategy allows us to fully capture environmental effects and galaxy-galaxy correlations while maintaining a low computational cost, thereby enabling the creation of more realistic mock galaxy catalogues. 

To generate a diverse and realistic population of galaxies, we adopt diffusion models \citep{Song20, Ho20}, which have recently demonstrated remarkable success in modelling complex high-dimensional data. Diffusion models have also been successfully applied to three-dimensional point cloud generation tasks \citep{Wang24_diffusion_model_pc_survey} and have been explored in astrophysical applications \citep{Cuesta-Lazaro24}.
Specifically, we use the Elucidated Diffusion Model (EDM) framework \citep{Karras22_EDM}, which reformulates diffusion models in terms of a continuous noise scale and provides improved training and sampling strategies compared to earlier discrete-time diffusion formulations.

In the following, we describe the dataset and model architecture in \cref{sec:methods} and show the results in \cref{sec:results}. We discuss future applications and possible extensions in \cref{sec:discussion}, before summarizing our conclusions in \cref{sec:conclusion}.
We adopt a flat cold dark matter model and $\Omega_{\rm m} = 0.3089$, $\Omega_{\rm  \Lambda} = 0.6911$, $h= 0.6774$, $\sigma_8 = 0.8159$, and $n_{\rm s} = 0.9667$ throughout this paper.

\section{Methods}
\label{sec:methods}

\subsection{Dataset}
We use the IllustrisTNG (The Next Generation; hereafter TNG) suite of cosmological hydrodynamic simulations \citep{Nelson19,Pillepich18b,Springel18,Nelson18a,Naiman18,Marinacci18} as our training dataset. 
The subgrid models in TNG are calibrated to reproduce several major observational results, including the cosmic star formation rate (SFR) history, stellar mass function, and stellar-to-halo mass relation at $z = 0$ \citep{Pillepich18a}. The simulation results are consistent with the observed galaxy properties including the galaxy clustering at $z < 2$ \citep{Springel18}, and some more comparisons with observations are made in, e.g. \citet{Springel18,Nelson18a,Naiman18,Marinacci18}.
In TNG, haloes are identified with a friends-of-friends (FoF) group finder, and within each halo, substructures, i.e., galaxies, are further identified using \textsc{SubFind} algorithm \citep{Springel01}. 

We use a snapshot at $z = 2$, which corresponds to the active star-formation epoch and a target epoch of upcoming galaxy surveys, from the TNG300 runs, which have a box size of $(302.6 \, \mathrm{Mpc})^3$.
To develop a model capable of generating small and faint galaxies even when provided with low-resolution DM distributions, we train the model to map low-resolution DM density fields to the galaxies in a much higher resolution.
More specifically, we degrade the TNG300-3-Dark data by randomly selecting one-tenth of the dark matter particles to construct the input DM density maps. This new data can be regarded as a proxy for an extremely low-resolution DM-only simulation.
Given the original particle mass resolution of $4.5\times 10^{9}~h^{-1} {\rm M_\odot}$, the downsampled data have a particle mass of $4.5\times 10^{10}~h^{-1}{\rm M_\odot}$. Assuming haloes containing at least $\sim 100$ particles can be reliably identified, this means that only haloes with masses $M_{\rm h} \gtrsim 4.5\times 10^{12}~h^{-1}{\rm M_\odot}$ can be detected in this data. 

From the downsampled TNG300-3-Dark snapshot containing $\sim 290^3$ DM particles, we construct a DM density fields with $256^3$ grids using the Cloud-In-Cell (CIC) mass assignment scheme. 
This corresponds to a voxel size of $(1.2~\rm Mpc)^3$. 
Ideally, the model should learn from as large a region as possible. However, simultaneously capturing both large-scale structures and small-scale features quickly becomes computationally prohibitive. Fortunately, galaxy formation is primarily governed by local physical conditions, and correlations between galaxy properties are typically limited to scales of a few Mpc.
Motivated by this, we design a model that generates galaxy point clouds from DM density fields on a patch-by-patch basis. We set the side length of a patch to $l_{\rm patch} = 18.9~\rm Mpc$. Each patch consists of $16^3$ voxels. 

To generate galaxies across a larger volume, we divide the simulation volume into non-overlapping patches that tile the entire domain, and then generate galaxies independently in each patch.
A potential concern with this approach is that discontinuities or artificial features may appear near patch boundaries, because the model has no explicit information about neighboring patches during generation. 
While this is problematic in map-based methods \citep[e.g.][]{Mishra26a}, we find that the simple tiling procedure is sufficient for our purpose. This is probably owing to the stochastic and sparse nature of star-forming galaxies.
We therefore present the results obtained with this procedure in the main text.
For completeness, we develop a network that incorporates galaxies outside the target region as conditioning information to promote the smoothness,
and present the details of our conditional method and its results in \cref{app:patch-wise-generation}.

As for target galaxies, we use those from TNG300-1, which has the particle resolution of $4.0\times 10^{7}~h^{-1}{\rm M_\odot}$ for DM and $7.6\times 10^{7}~h^{-1}{\rm M_\odot}$ for baryon, resolving down to $M_{\rm h} \sim 4 \times 10^{9}~h^{-1}{\rm M_\odot}$ haloes. 
We train the model to generate galaxies with ${\rm SFR} > 1~\rm M_\odot/yr$. 
This criterion is sufficient for constructing mock catalogues for galaxy redshift surveys targetted to emission line galaxies, but also for capturing the bulk of the galaxy population relevant for LIM analyses.\footnote{
For instance, ${\rm SFR} = 1~\rm M_\odot/yr$ corresponds to an H$\rm \alpha$ luminosity of $L_{\rm H\alpha} = 10^{41}~\rm erg/s$, which gives fluxes of $f_{\rm H\alpha} = 2\times 10^{-17}~\rm erg/s/cm^2$ at $z = 1$ and $4\times10^{-18}~\rm erg/s/cm^2$ at $z = 2$. 
These fluxes are below the nominal line-flux limit quoted for Euclid ($2\times 10^{-16}$; \citealt{Euclid22}).
For LIM, galaxies above this threshold capture most of the signal, contributing approximately 90 percent of the total SFR at $z = 2$ in the whole TNG volume.
}
With this selection and the voxel resolution of $(1.2~\rm Mpc)$, we find that a non-negligible number of voxels are occupied by multiple galaxies, suggesting that a map-to-map modelling framework is not sufficient to capture the individual galaxies.

The feature vector of each galaxy is denoted by $\bm{s}$, a five-dimensional vector consisting of its three-dimensional position, SFR, and stellar mass.
Within each patch, the galaxy positions are normalized such that the patch range corresponds to $[-1,1]$ along each spatial axis. 
The SFR and $M_\star$ values are normalized as 
\begin{align}
    2 \, \frac{\log s - a}{b - a} - 1,
\end{align}
with $(a, b)$ being set to $(0,3)$ for SFR (in $\rm M_\odot/yr$) and $(7, 12)$ for $M_\star$ (in $\rm M_\odot$). 
This transform maps the typical dynamic ranges of these quantities to approximately $[-1,1]$. 
The input DM density $\rho$ (in ${\rm M_\odot / (kpc}/h)^3$) is normalized as $\log \rho / 5$, which transforms the values to approximately fall within the range $[0,1]$.

For each patch, we set the maximum number of galaxies to $L = 200$.
This value is sufficiently larger than the maximum number of target galaxies in any patch in the dataset.
If a patch contains fewer than $L$ galaxies, we pad the remaining entries with dummy galaxies.
To distinguish between actual and null galaxies, we extend the feature vector $\bm{s}$ with an additional flag variable, taking the value 1 for real galaxies and $-1$ for null (padded) samples.

\subsection{Diffusion Model}

In the EDM formulation, a clean target data point $x_0$ is perturbed by additive Gaussian noise with varying noise levels $\sigma$, producing a noisy sample $x_\sigma$:
\begin{align} \label{eq:x_sigma}
x_\sigma = x_0 + \sigma\,\epsilon, \quad \epsilon \sim \mathcal{N}(0, I),
\end{align}
where $\sigma$ is a continuous noise scale sampled from a predefined distribution $p(\sigma)$.

During training, a neural network $D_\theta(x_\sigma, \sigma, y)$ is trained to predict the clean data $x_0$ from the noisy input.
Here, $\theta$ denotes the trainable parameters of the network.
Following \citet{Karras22_EDM}, noise-dependent input and output scalings are used to stabilize training across a wide range of noise levels, and a weighted denoising objective is adopted:
\begin{align}
\mathcal{L}_{\rm EDM} = \mathbb{E}_{x_0, \epsilon, \sigma} \left[ w(\sigma) \left| D_\theta(x_\sigma, \sigma, y) - x_0 \right|^2 \right],
\end{align}
where $w(\sigma)$ is a noise-dependent weighting function designed to balance contributions across different noise levels.
We follow the standard EDM parameterization and preconditioning scheme to improve numerical stability.

Given a denoiser $D_\theta(x,\sigma)$ that estimates the underlying clean
sample $x_0$ from a noisy input $x$ at noise level $\sigma$, the corresponding
EDM probability-flow ordinary differential equation can be written as
\begin{align}
    \label{eq:ODE}
    \frac{dx}{d\sigma}
    =
    \frac{x - D_\theta(x,\sigma)}{\sigma}.
\end{align}
This can be obtained by replacing the unkown $x_0$ with the prediction $D_\theta(x,\sigma)$ and setting $d x_\sigma / d\sigma = \epsilon$ in \cref{eq:x_sigma}. 
At inference time, generation starts from a highly noisy sample $x_{\sigma_{\max}} \sim \mathcal{N}(0,\sigma_{\max}^2 I)$.
The sample is then progressively denoised by moving along a sequence of decreasing noise levels,
$\sigma_{\max} = \sigma_0 > \sigma_1 > \cdots > \sigma_{N-1} = \sigma_{\min}$. 
At each noise level $\sigma_i$, the denoiser predicts a clean estimate $D_\theta(\hat{x}_{\sigma_i},\sigma_i) \approx x_0$ from the current noisy sample $\hat{x}_{\sigma_i}$.
This estimate defines the direction in which the sample should be updated toward a lower-noise state.
We use the second-order Heun method for sampling.

An important practical issue is the choice of the sampling schedule, which controls the trade-off between generation quality and computational cost.
While previous study has discussed suitable choices for general image-generation tasks \citep{Karras22_EDM}, the optimal choice may depend on the data distribution and the desired computational efficiency.
We therefore examine this issue for our data in \cref{app:sampling_schedule}. 
We find that the generation quality is already stable for $\gtrsim 20$ sampling steps. 
With this number of steps and the network implementation described below, generating the full $(151.3~\rm Mpc)^3$ test volume takes approximately 10 seconds in our computational environment.

\subsection{Network}

For the denoising network, we develop a transformer-based architecture designed to model point-wise features conditioned on a data cube. 
Transformers \citep{Vaswani17} are well suited to our task as self-attention allows the model to learn interactions among galaxies represented as an unordered set of points.
Following the EDM convention, the network takes as input a set of points $x_\sigma \in \mathbb{R}^{L\times 5}$, a noise-level embedding $c_{\rm noise} = \frac{1}{4}\log \sigma$, and conditioning variable $y$, and outputs $\epsilon_\theta (x_\sigma, t, y)$. 
The condition $y$ is a three-dimensional DM density cube. 

The DM density cube is encoded by a voxel encoder based on a three-dimensional Vision Transformer \citep{Dosovitskiy20}, using patch tokens obtained by a 3D convolution with kernel size and stride of 2, followed by a three-layer transformer encoder and LayerNorm.
The input cube is first mapped to patch tokens using a 3D convolution with kernel size 2 and stride 2. After adding positional embeddings, the token sequence is processed by a three-layer transformer encoder, followed by LayerNorm.
The noisy input points $x_\sigma$ are first projected into the latent space with a two-layer multilayer perceptron MLP and treated as query tokens. At each decoder block, point features are updated through self-attention, followed by cross-attention to the conditioning sequence, and a point-wise feed-forward network. 
The noise-level embedding $c_{\rm noise}$ is first mapped to a sinusoidal embedding \citep{Vaswani17} and then injected into every decoder layer through adaptive layer normalization (AdaLN) \citep{Peebles22}.
After four layers of self-/cross-attention, the final point-wise hidden features are projected back to the original point dimension to predict the noise term $\bm{\epsilon}_\theta$.
We adopt the hidden dimension $d_{\rm model} = 128$.

\subsection{Training}

We reserve a cubic volume of $(151.3~\rm Mpc)^3$ for test and use the rest of the volume for training and validation. For training and validation data, we randomly sample 90\,000 and 10\,000 patches, respectively. 
We trained the model with the batch size of 512 with the Adam optimizer \citep{Kingma14} and a learning rate of $10^{-4}$ and a weight decay of $10^{-3}$. 
The model is implemented in \texttt{PyTorch} \citep{pytorch}, and both training and generation are performed on a single NVIDIA RTX 6000 Ada Generation GPU. 
The model is trained for 10,000 epochs.

\section{Results}
\label{sec:results}

\subsection{Visual Inspection}

\begin{figure}
    \centering
        \includegraphics[width=8cm]{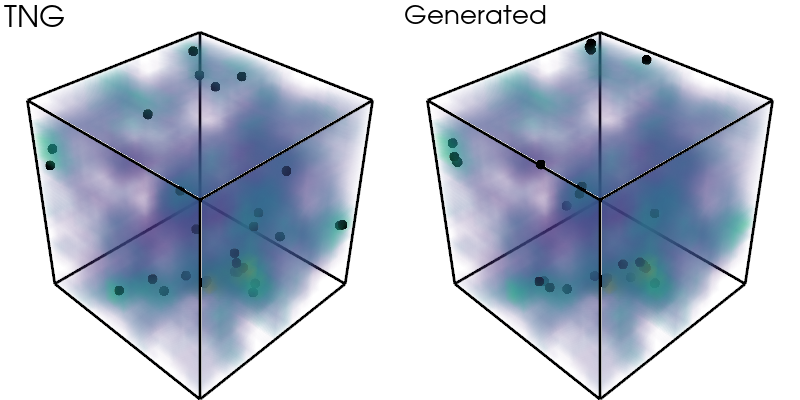}
        \includegraphics[width=8cm]{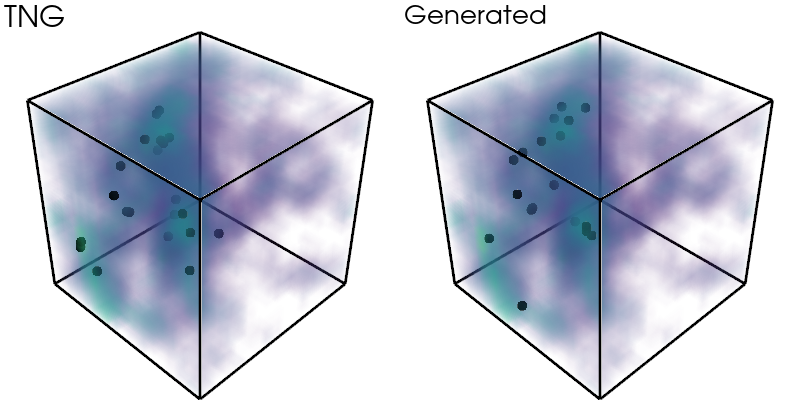}
        \includegraphics[width=8cm]{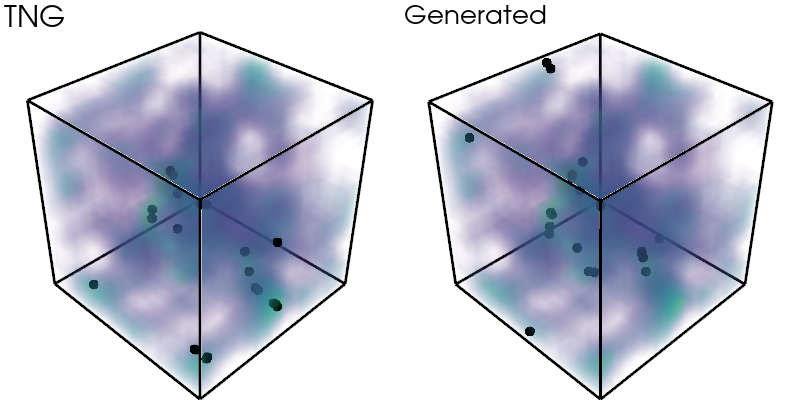}
        \includegraphics[width=8cm]{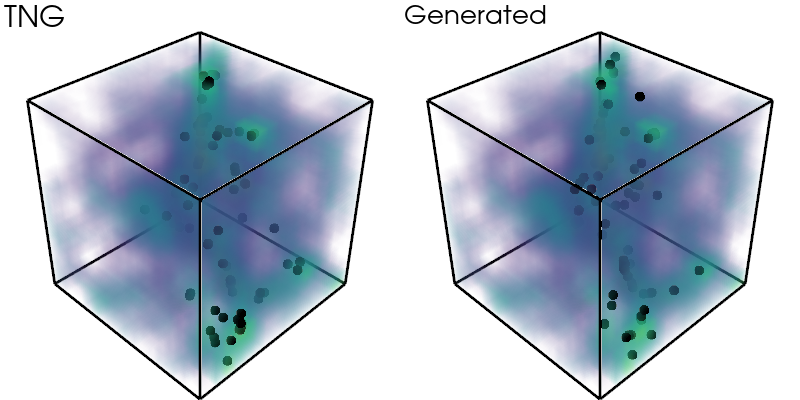}
    \caption{True and generated point distributions in patches randomly extracted from the test region. Each patch has a side length of 18.9 Mpc. 
    }
    \label{fig:volumes}
\end{figure}

\Cref{fig:volumes} shows the true and generated point distributions in four randomly selected test patch cubes.
The colormap represents the underlying DM density field used as the conditioning input.
In all examples, the generated galaxies preferentially populate the high-density regions and are distributed along the filamentary DM structures.
This visual agreement indicates that the model has learned the spatial correspondence between the continuous DM density field and the discrete galaxy point process at the patch level.
We note that the generated catalogue is not expected to reproduce the exact galaxy positions in the TNG catalogue on an object-by-object basis, because star formation activity has a stochastic feature and the diffusion model samples from a conditional distribution.
The relevant comparison is therefore whether the generated samples reproduce the {\it statistical} properties of the galaxy distribution conditioned on the same DM field.

\begin{figure*}
    \centering
        \includegraphics[width=16cm]{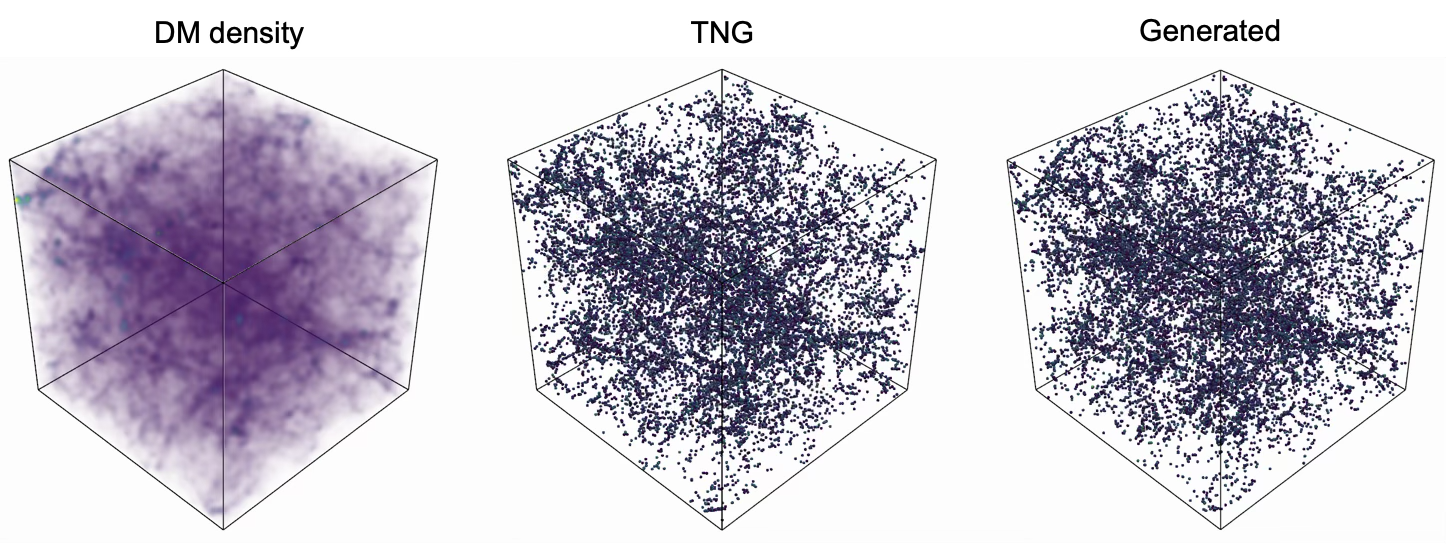}
    \caption{DM density (left), true TNG galaxies (middle), and generated galaxies (right) in the test region of a volume of $(151.3 ~\rm Mpc)^3$}
    \label{fig:distribution_3d}
\end{figure*}

We then apply the trained model to the full test region by dividing it into non-overlapping patch cubes, generating galaxies in each patch independently, and concatenating the resulting catalogues. 
A very small fraction of generated galaxies fall outside their nominal patch boundaries, but these objects are retained when constructing the full-volume catalogue.
\Cref{fig:distribution_3d} shows the conditioning DM density field (left), the true TNG galaxy distribution (middle), and the generated galaxy distribution (right) over the entire test region.
All galaxies with ${\rm SFR} > 1~{\rm M_\odot /yr}$ are shown.
The generated catalogue traces the same large-scale cosmic web as the TNG galaxies, but individual small-scale configurations are not identical. 
This visual comparison suggests that the patch-level outputs can be assembled over the full test volume while preserving the global spatial organization of galaxies.
We quantitatively examine the reproduction of large-scale clustering in the next section using summary statistics.

\subsection{Summary Statistics}
\begin{figure*}
    \centering
        \includegraphics[width=16cm]{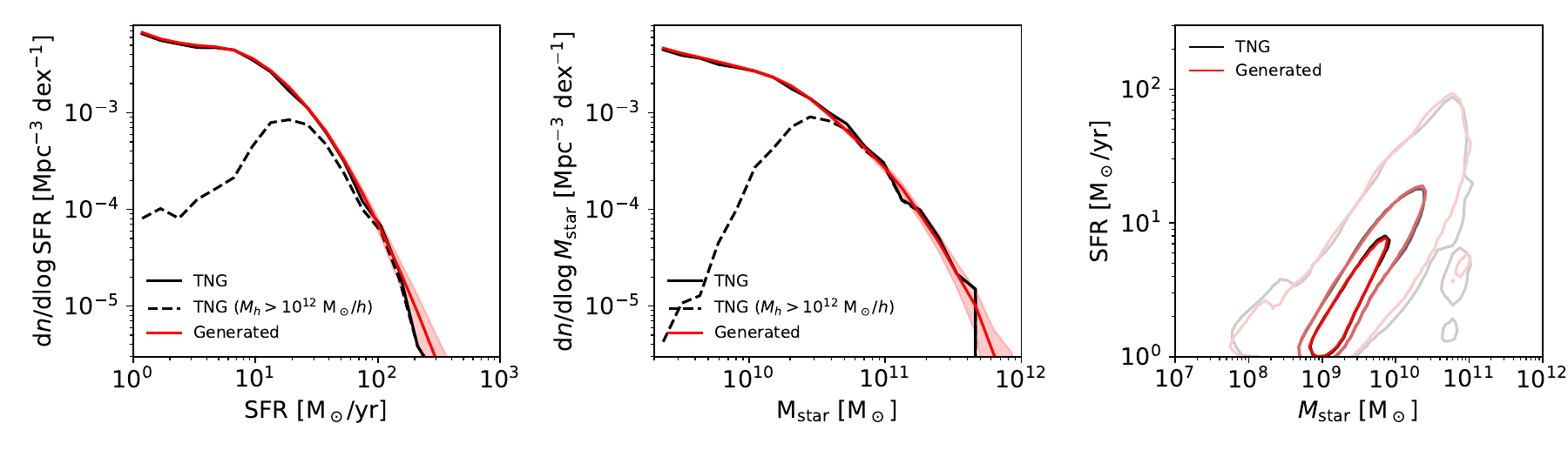}
    \caption{Histograms of SFRs (left) and stellar masses (middle) of TNG (black) and generated galaxies (red). 
    For the generated ones, the shaded regions show the variance of 100 different realizations. For comparison, the black dashed line show that of TNG galaxies with halo mass greater than $10^{12}~{\rm M_\odot}/h$. Right panel shows the two-dimensional histogram. }
    \label{fig:hist}
\end{figure*}

\Cref{fig:hist} compares the distributions of galaxy properties in the TNG catalogue (black) and the generated catalogues (red).
The left and middle panels show the one-dimensional distributions of SFR and stellar mass, respectively, while the right panel shows their joint distribution.
The generated samples reproduce both the marginal distributions of SFR and stellar mass and their correlation in the two-dimensional plane.
For the generated catalogues, the shaded regions show the standard deviation of 100 independent realizations.
Although individual patches can exhibit more diversity across realizations, these variations are largely averaged out when the statistics are measured over the full volume.

To highlight the advantage of our approach, which directly conditions on the DM density field, it is useful to compare it with a halo-based approach that also starts from the underlying matter distribution. In such an approach, one would first identify haloes in the DM-only simulation corresponding to the input density field, and then populate them with galaxies using a halo-galaxy connection model, either based on machine learning or on more conventional empirical prescriptions.
The black dashed line in \cref{fig:hist} shows the distribution of TNG galaxies hosted by {\sc Subfind} subhaloes with masses above $10^{12}~h^{-1}{\rm M_\odot}$, which approximately corresponds to the mass of haloes resolved with 20 particles in the low-resolution data used as an input in this study.
This represents the idealized limit of a resolved-halo-based approach with a perfect halo-galaxy connection model. 
The comparison shows that even in such this idealized case, such an approach would miss a substantial population of galaxies with small SFRs and stellar masses, whose host haloes fall below the nominal resolution limit.
In contrast, our diffusion-based method reproduces a galaxy population that extends well below this limit.
This is because the model effectively marginalizes over the unresolved small-scale structures that are absent from the input, allowing galaxies associated with these structures to be generated statistically from the low-resolution (or smoothed) DM density field.

\begin{figure}
    \centering
        \includegraphics[width=7.5cm]{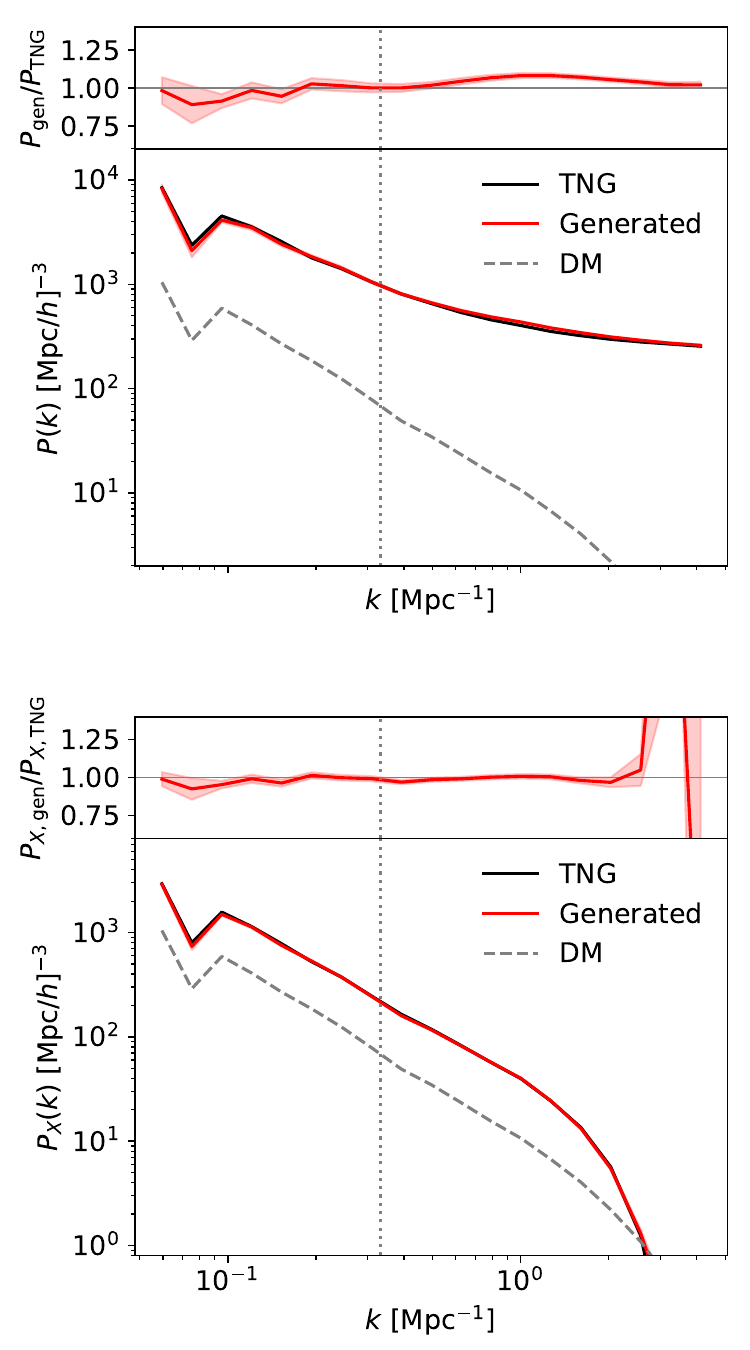}
    \caption{Top: Power spectrum of galaxy number density map. For both TNG and generated catalog, galaxies with ${\rm SFR} >1 ~\rm M_\odot/yr$ are used. The vertical dotted line indicates the patch-size scale $k_{\rm patch} = 2\pi / l_{\rm patch}$. Bottom: Cross-power spectrum between galaxy number density map and DM density map. The upper panels show the ratios of the predicted power spectra to the corresponding TNG results.}
    \label{fig:power}
\end{figure}

We further examine the spatial clustering of the generated galaxies.
\Cref{fig:power} shows the auto-power spectrum of the galaxy number-density field (top) and the cross-power spectrum between galaxies and the DM density field (bottom). 
The dashed curves show the corresponding DM power spectrum for reference. 
The vertical dotted line indicates the scale corresponding to the patch size, $k_{\rm patch}=2\pi/l_{\rm patch}$. 
The generated catalogues reproduce the TNG galaxy auto-power spectrum over a wide range of scales. 
The agreement in the galaxy-DM cross-power spectrum further shows that the generated galaxies are correctly correlated with the underlying matter distribution.
The small-scale agreement reflects the fact that the generated point clouds closely follow the local DM density structures within individual patches, as also illustrated in \Cref{fig:volumes}.
The absence of noticeable discrepancies around the patch scale and at smaller scales also suggests that patch assembly does not introduce significant boundary artifacts.
On large scales, the agreement can be understood as a consequence of the model learning how the average galaxy abundance varies from patch to patch in response to the conditioning DM density field, thereby reproducing the effective galaxy bias.
The lack of a significant large-scale discrepancy further shows that, at least for the two-point statistics considered here, the density structure encoded within each input patch is sufficient; explicit information from neighbouring patches is not required to reproduce these statistics.
We note, however, that agreement in the auto- and cross-power spectra does not by itself guarantee the reproduction of higher-order correlations or mode-mode coupling. If such statistics are important for a particular application, they should be validated separately.

\begin{figure}
    \centering
        \includegraphics[width=7.5cm]{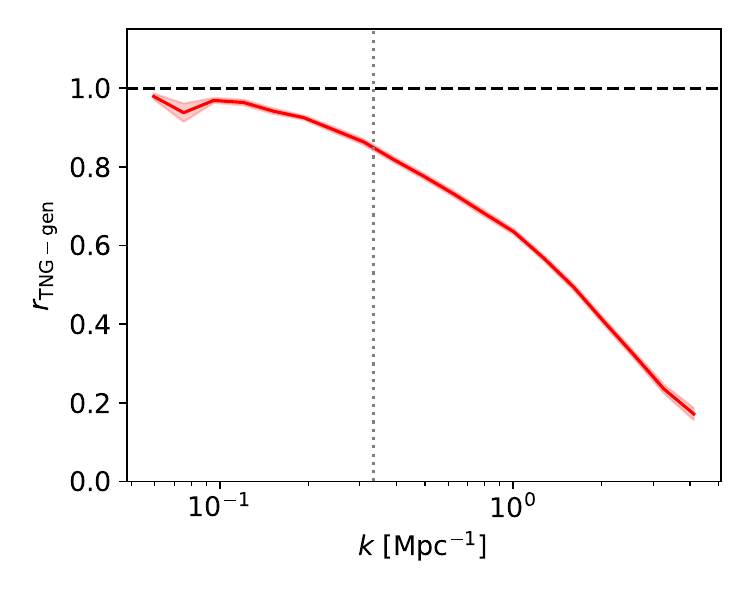}
    \caption{Cross-correlation coefficient between the TNG and generated of galaxy distributions. For both TNG and generated catalog, galaxies with ${\rm SFR} >1 ~\rm M_\odot/yr$ are used.}
    \label{fig:r_true-gen}
\end{figure}

Finally, to quantify the spatial correspondence between the TNG and generated distributions, we compute the cross-correlation coefficient, 
\begin{align}
    r_{\rm TNG-gen}(k) = \frac{P_{\rm TNG-gen}(k)}{\sqrt{P_{\rm TNG}(k)\, P_{\rm gen}(k)}}, 
\end{align}    
where $P_{\rm TNG-gen}(k)$ is their cross-power spectrum and $P_{\rm TNG}(k)$ and $P_{\rm gen}(k)$ are the respective auto-power spectra. 
\Cref{fig:r_true-gen} shows the computed correlation coefficient.
The generated distribution remains strongly correlated with the TNG distribution on large scales, with $r_{\rm TNG-gen}$ close to unity, demonstrating that the model correctly captures the large-scale structures determined by the common dark matter field. 
The correlation gradually decreases toward smaller scales (higher $k$).
Nevertheless, $r_{\rm TNG-gen}$ remains above zero even at high $k$, indicating that the model retains substantial spatial information from the conditioning dark matter field down to small scales. 
At the same time, its departure from unity reflects the stochastic component of galaxy formation that is not fully specified by the conditioning DM density field. 
Importantly, our goal is not to reproduce the particular TNG realization galaxy by galaxy, but to generate statistically consistent stochastic realizations of the galaxy distribution conditioned on the dark matter field. 
The reduced but non-zero small-scale correlation therefore indicates a combination of physically meaningful spatial correspondence and the intended probabilistic generative behaviour.

\section{Extensions and Scalability}
\label{sec:discussion}

The results presented above demonstrate the feasibility of generating galaxy catalogues with a diffusion-based model conditioned on dark-matter density fields. This framework can be extended in several directions.
A natural extension is to train the model on a broader range of redshifts and simulations, which would allow the generated catalogues across different cosmic epochs, galaxy populations, and cosmological or astrophysical models.
Such information can be incorporated as global conditioning variables. For example, redshift or simulation parameters can be embedded with a small neural network and used to modulate the denoising network, through additional conditioning tokens or adaptive normalization. 

Another important extension is to increase the set of galaxy properties generated by the model.
While the present implementation focuses on positions, stellar masses, and SFRs, the same formulation can in principle be applied to other physical properties and observables, such as velocities, gas masses, and metallicities. 
This would allow the model to generate mock catalogues that jointly predict multiple galaxy populations and observables relevant to different surveys. 
Importantly, adding such properties is computationally straightforward in the present point-cloud diffusion framework. 
Additional quantities can be included by increasing the dimensionality of the feature vector associated with each galaxy, without increasing the number of galaxy tokens processed by the transformer. 
This is in contrast to token-by-token autoregressive approaches \citep[e.g.][]{Moriwaki26}, where modelling multiple correlated properties may require representing them as additional tokens and therefore increasing the effective sequence length.
Since the computational cost of transformer-based models is typically more sensitive to the number of tokens than to the dimensionality of each token, extending the feature vector is not expected to be a major computational bottleneck compared with increasing the number of galaxies per patch.

Achieving high-precision predictions for a wider range of observables at a reasonable computational cost may require identifying the spatial extent of the dark-matter field that is most relevant for modelling the galaxy population. 
While the statistics examined in this study are well reproduced with the adopted patch size, the optimal patch size remains to be explored: larger patches may be beneficial for more demanding observables or accuracy requirements, whereas smaller patches may provide comparable performance at reduced cost. 
Training the model with different conditioning volumes and applying it to large simulation boxes would provide a direct way to quantify this dependence across different summary statistics. 

A practical limitation is that increasing the patch size at fixed spatial resolution rapidly increases the number of mesh elements processed by the transformer-based encoder, leading to higher memory and computational requirements. 
In our implementation and computational environment, a conditioning field of size $\sim 32^3$ was close to the practical limit. Applications involving larger input fields may therefore benefit from more efficient architectures, such as latent diffusion models \citep{Rombach21_LDM}, more scalable encoders \citep[e.g.][]{Liu21_SwinTransformer}, or sparse-attention variants \citep[e.g.][]{Beltagy20_Longformer, Zaheer20_BigBird}.

\section{Conclusions}
\label{sec:conclusion}

We have presented a diffusion-based method for generating galaxy mock catalogues from dark matter density fields.
In our approach, the dark matter field is provided as a three-dimensional conditioning mesh, while galaxies are generated directly as a point cloud with positions and physical properties.
We trained the model on the IllustrisTNG simulation using the Elucidated Diffusion Model framework and applied it to generate galaxies with ${\rm SFR} > 1~{\rm M_\odot /yr}$ in a test volume.

The generated catalogues reproduce several key properties of the TNG galaxy distribution.
They trace the filamentary structures of the underlying dark matter field, recover the one- and two-dimensional distributions of SFR and stellar mass, and reproduce the galaxy auto-power spectrum and the galaxy-DM cross-power spectrum.
We also showed that the method can generate a galaxy population extending below the nominal halo-resolution limit of the input dark matter field.
This is possible because the model effectively marginalises over unresolved small-scale structures, allowing galaxies associated with such structures to be generated statistically from the low-resolution density field.

With an improved sampling strategy (\cref{app:sampling_schedule}), the full catalogue in the $(151.3~\rm Mpc)^3$ volume can be generated in approximately 10 seconds on a single GPU.
Since different spatial regions can be generated independently, the method is naturally parallelizable and can, in principle, be applied to larger volumes by dividing them into patches.
Moreover, because galaxies are represented as discrete points rather than voxelized fields, the impact of patch boundaries is less severe than in map-to-map generation methods (see \cref{app:patch-wise-generation} for discussion).
The point-cloud representation also makes the output flexible: galaxy positions and properties are generated as continuous quantities rather than being tied to a fixed output grid, allowing the catalogues to be post-processed for different survey geometries, angular or  

The framework is also extensible. Additional galaxy properties or observables can be included by increasing the dimensionality of the per-galaxy feature vector, without changing the number of galaxy tokens processed by the model. This makes it straightforward to generate richer catalogues, enabling predictions for a wide range of observational targets. 
The conditioning information can also be extended to include redshift or simulation parameters, enabling applications across different cosmic epochs or galaxy-formation models.

Our results demonstrate that conditional diffusion models provide a promising route toward fast and flexible mock generation from low-resolution matter fields, complementing conventional halo-based approaches. Future study will extend this framework to a broader range of galaxy populations, physical properties, and summary statistics, and will further investigate the scales of the dark matter field required to reproduce different observables at high precision.

\section*{Acknowledgements}
This work was supported in part by Japan Society for the Promotion of Science (JSPS) KAKENHI Grant Numbers JP23K03446, JP23K20035, JP24H00004 (K.~M.), 
JP24H00215, JP25H00662, JP25H01513, and JP25K17380 (K.~O.).

\section*{Data Availability}
The TNG simulation data used for training is publicly available.\footnote{https://www.tng-project.org}
We will share the scripts and the trained model used in this paper upon
reasonable request.

\bibliographystyle{mnras}
\bibliography{bibtex_library} 


\appendix
\crefalias{section}{appendix}

\section{Autoregressive patch-based generation}
\label{app:patch-wise-generation}

\begin{figure*}
  \centering
  \begin{minipage}{0.65\textwidth}
    \centering
    \includegraphics[width=\textwidth]{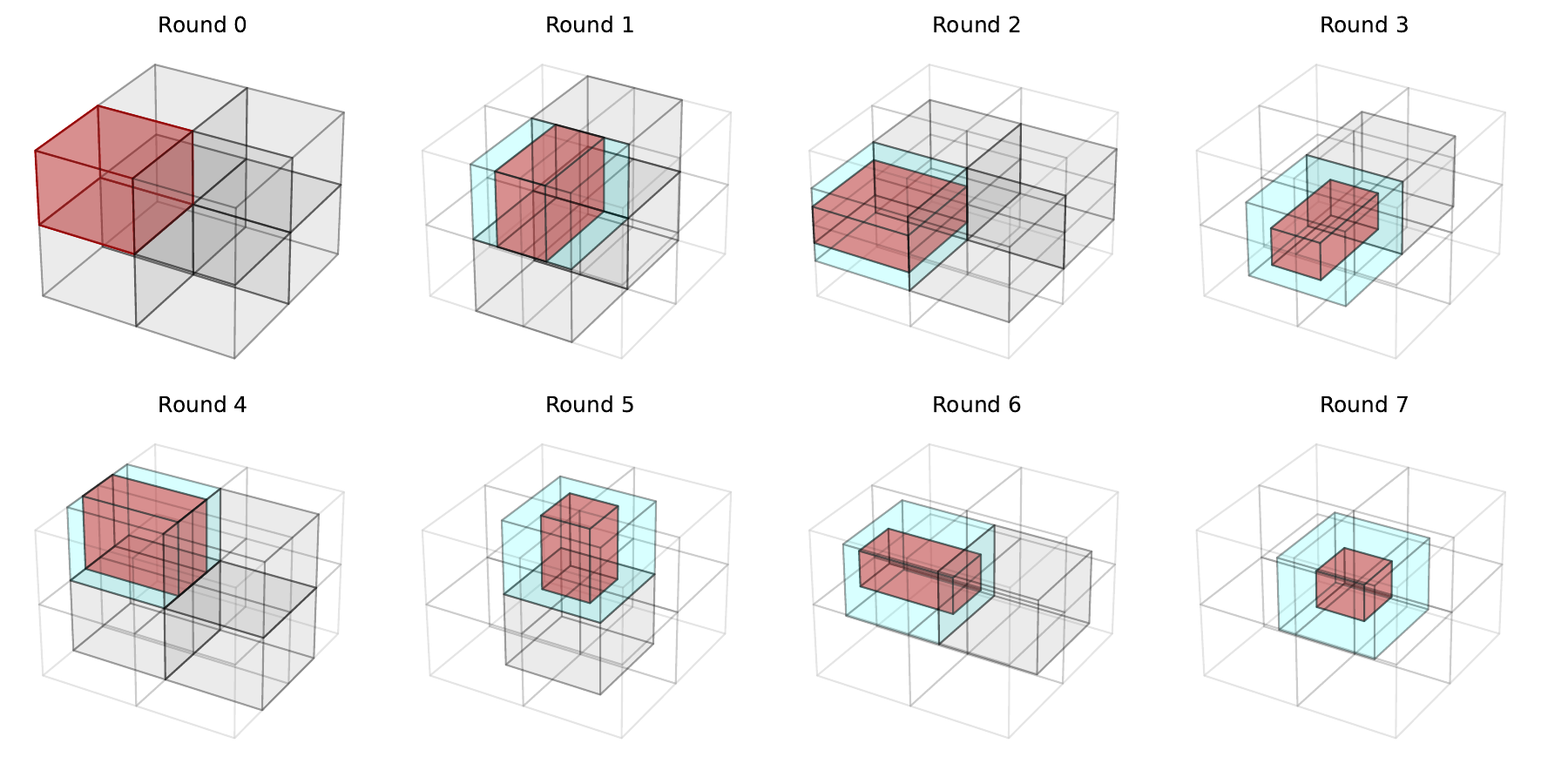}
  \end{minipage}
  \begin{minipage}{0.3\textwidth}
    \centering
    \includegraphics[width=\textwidth]{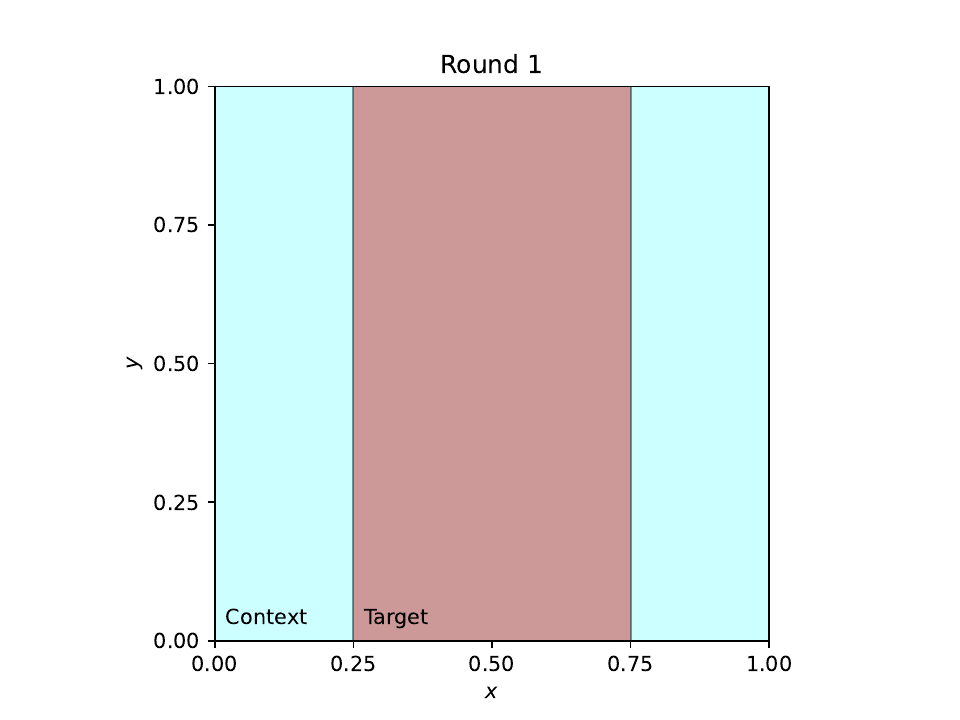}
  \end{minipage}
  \caption{Autoregressive generation process. In each round, the data domain is covered using a different patch configuration and a different assignment of target and context regions. Galaxies generated in early rounds are replaced by the newly generated ones if they lie within the target regions of the current round.}
  \label{fig:bboxes}
\end{figure*}

If one simply applies the model independently to non-overlapping patches when generating a larger volume, the continuity of the galaxy distribution across patch boundaries is not guaranteed.
A common way to address this problem is to use an autoregressive scheme, in which galaxies generated in previous steps are included as conditioning information for later steps (e.g. \citealt{Rouhiainen24,Mishra26a}; see also \citealt{Shirasaki24_neural_style_transfer,Mishra26b} for alternative strategies).
We adopt a similar approach here.
For computational efficiency, we organize the generation procedure into eight rounds, as illustrated in \cref{fig:bboxes}. 
In each round, galaxies that have already been generated and fall within the context regions (cyan region) are used as conditioning information to generate new galaxies in the target regions (red regions). The newly generated galaxies then overwrite any galaxies that had previously been generated in the same target regions and are used in subsequent rounds.
This construction is designed so that no galaxy is generated in isolation; If the model successfully learns the relevant spatial correlations, the generation of any given galaxy can depend on neighboring galaxies within at least $\sim 0.25,l_{\rm patch}$, either through joint generation or through conditioning on the context region.
Because the patches in the same round are non-overlapping and therefore can be generated independently, this framework can be more efficiently parallelized than commonly adopted raster-scan method \citep[e.g.][]{Rouhiainen24}.

To enable this procedure, the model is trained under eight different combinations of context and target regions. 
Let us denote the boundary box of the target region by $\bm{b} = (x_{\rm min}, x_{\rm max}, y_{\rm min}, y_{\rm max}, z_{\rm min}, z_{\rm max})$. 
We use the context galaxies $\bm{x}_{\rm ctx}$ and the boundary $\bm{b}$ in addition to the DM density cube as the condition $y$. 
The context galaxies are passed through a context encoder, where the properties of each galaxy are embedded by a two-layer multilayer MLP and further refined by a three-layer transformer encoder with a padding mask. 
The bounding coordinate $\bm{b}$ is mapped to the same latent dimension using a separate MLP.
These two components are concatenated with the voxel tokens into a single conditioning sequence.
To distinguish the source of each conditioning token, we add a learned type embedding indicating whether a token originates from the voxel, context point, or the boundary vector. 
The {\it null} context galaxies are masked out when fed into the transformer.

\begin{figure}
    \centering
        \includegraphics[width=8cm]{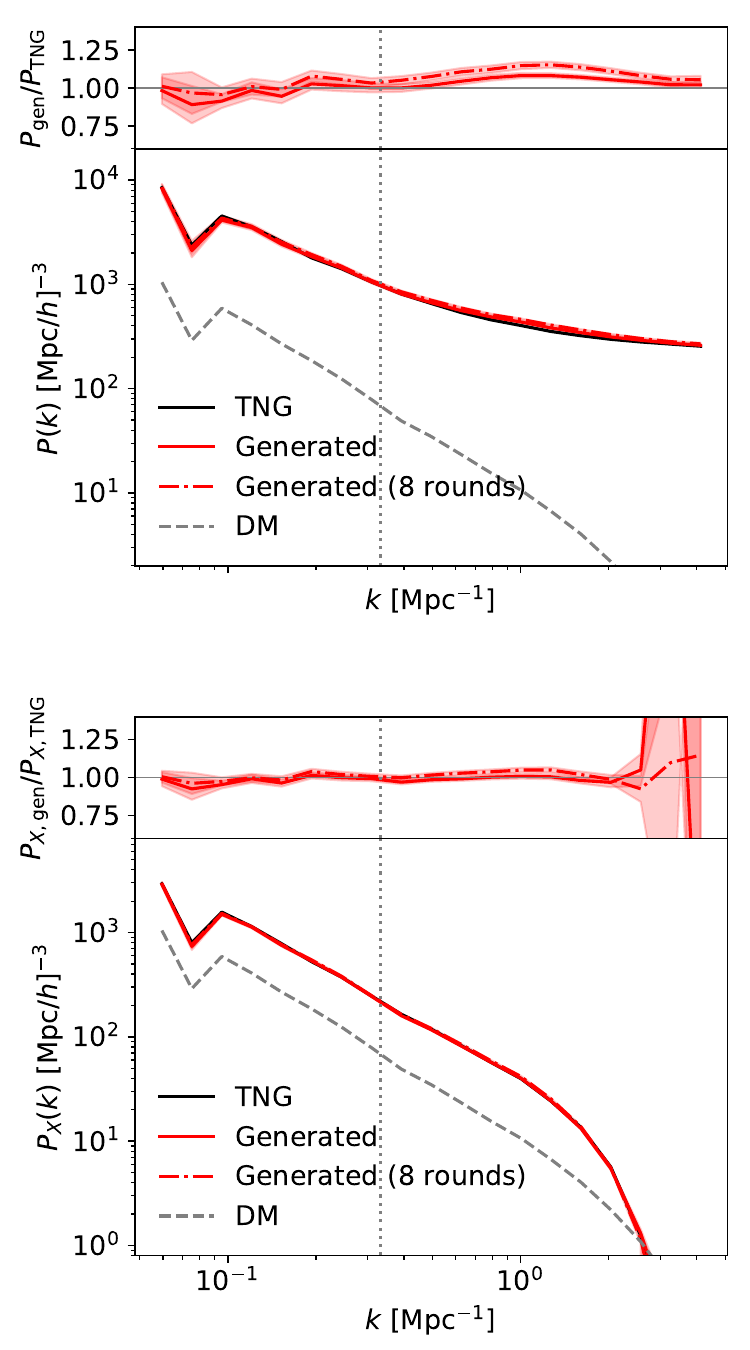}
    \caption{Same as \cref{fig:power}, with the dash-dotted lines additionally showing the power spectra measured from catalogues generated using the eight-round autoregressive scheme.
}
    \label{fig:power_8rounds}
\end{figure}

The power spectra obtained with the eight-round autoregressive generation scheme are shown by the dash-dotted lines in \cref{fig:power_8rounds}. 
Overall, they reproduce the TNG power to a similar accuracy as the one-round generation shown in \cref{fig:power}, and the corresponding curves almost overlap.

\section{Choice of sampling schedule}
\label{app:sampling_schedule}

The choice of the sampling schedule is an important practical issue since it controls the trade-off between generation quality and computational cost.
In the original EDM paper, \citet{Karras22_EDM} adopt a sampling schedule of the form
\begin{align} \label{eq:sampling}
    \sigma_i = \left( \sigma_{\max}^{1/\rho} + r_i 
    \left( \sigma_{\min}^{1/\rho} - \sigma_{\max}^{1/\rho} \right)
    \right)^\rho,
\end{align}
where
\begin{align}
    r_i = \frac{i}{N-1},
\end{align}
for $i < N$ and $\sigma_N = 0$, where the parameter $\rho$ is chosen empirically based on what performs best. A schedule with a larger $\rho$ samples the larger $\sigma$ region more finely. We find that increasing $\rho$ tends to improve the generation quality, consistent with the trend reported in \citet{Karras22_EDM}, but we do not observe a substantial difference for values larger than $\rho=7$. We therefore adopt $\rho=7$, the default value used in \citet{Karras22_EDM}, throughout this work.

\begin{figure}
    \centering
        \includegraphics[width=9cm]{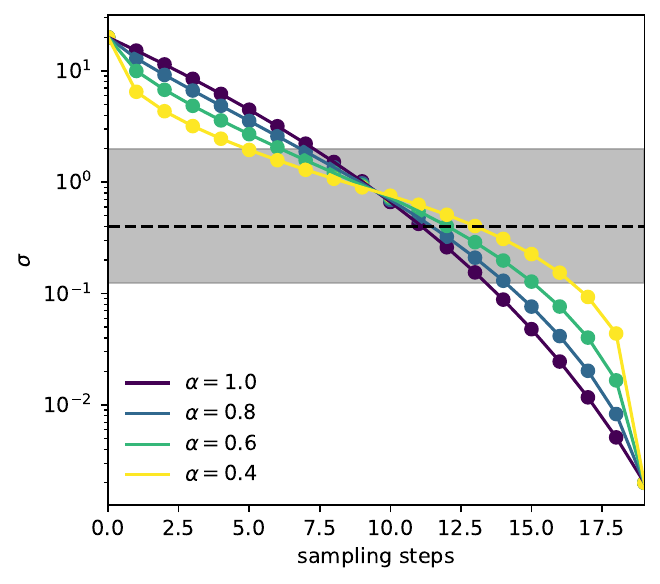}
    \caption{Sampling schedule with different sampling parameter $\alpha$ (see \cref{eq:sampling} and \cref{eq:r_i}). The upper boundary of the shaded region corresponds to the full dynamic range of the data. The lower boundary indicates the value corresponding to the pixel size of 1.2 Mpc. The horizontal dotted line marks the standard deviation of the normalized data, $\sigma \sim 0.4$. Sampling with $\alpha < 1$ put more focus on the intermediate scale.
    }
    \label{fig:sigma_schedule}
\end{figure}

Naively thinking, it may be beneficial to place more sampling points around the noise range corresponding to the effective dynamic range of the data, where the transition from noise-dominated states to data-dominated structures occurs. 
This region is likely to be important for accurately forming the relevant spatial structures while avoiding unnecessary steps in noise ranges where the reverse dynamics are comparatively smooth or less informative. 
To account for this, we consider a modified noise schedule with a denser discretization around the data-relevant noise range, while keeping fewer steps in less sensitive regions. We replace $r_i$ in \cref{eq:sampling} with
\begin{align} \label{eq:r_i}
r_i = 
\begin{cases} 
    \dfrac{1}{2}\left(\frac{2i}{N-1}\right)^\alpha, & i < \frac{N-1}{2}, \\[8pt]
    1 - \dfrac{1}{2}\left(\frac{2(1-i)}{N-1}\right)^\alpha, & i \geq \frac{N-1}{2}.
\end{cases}
\end{align}
with an additional free parameter $\alpha$. Sampling with $\alpha < 1$ put more focus on the intermediate scale. 
\Cref{fig:sigma_schedule} shows the sampling schedules with $\rho = 7$ and $\alpha = 1.0, 0.8, 0.6, 0.4$. 
The upper boundary of the shaded region corresponds to the full dynamic range of the data, which is 2 because the data are scaled to [-1,1]. The lower boundary indicates the value corresponding to the pixel size of 1.2 Mpc. The horizontal dotted line marks the standard deviation of the normalized data, $\sigma \sim 0.4$.

\begin{figure}
    \centering
        \includegraphics[width=9cm]{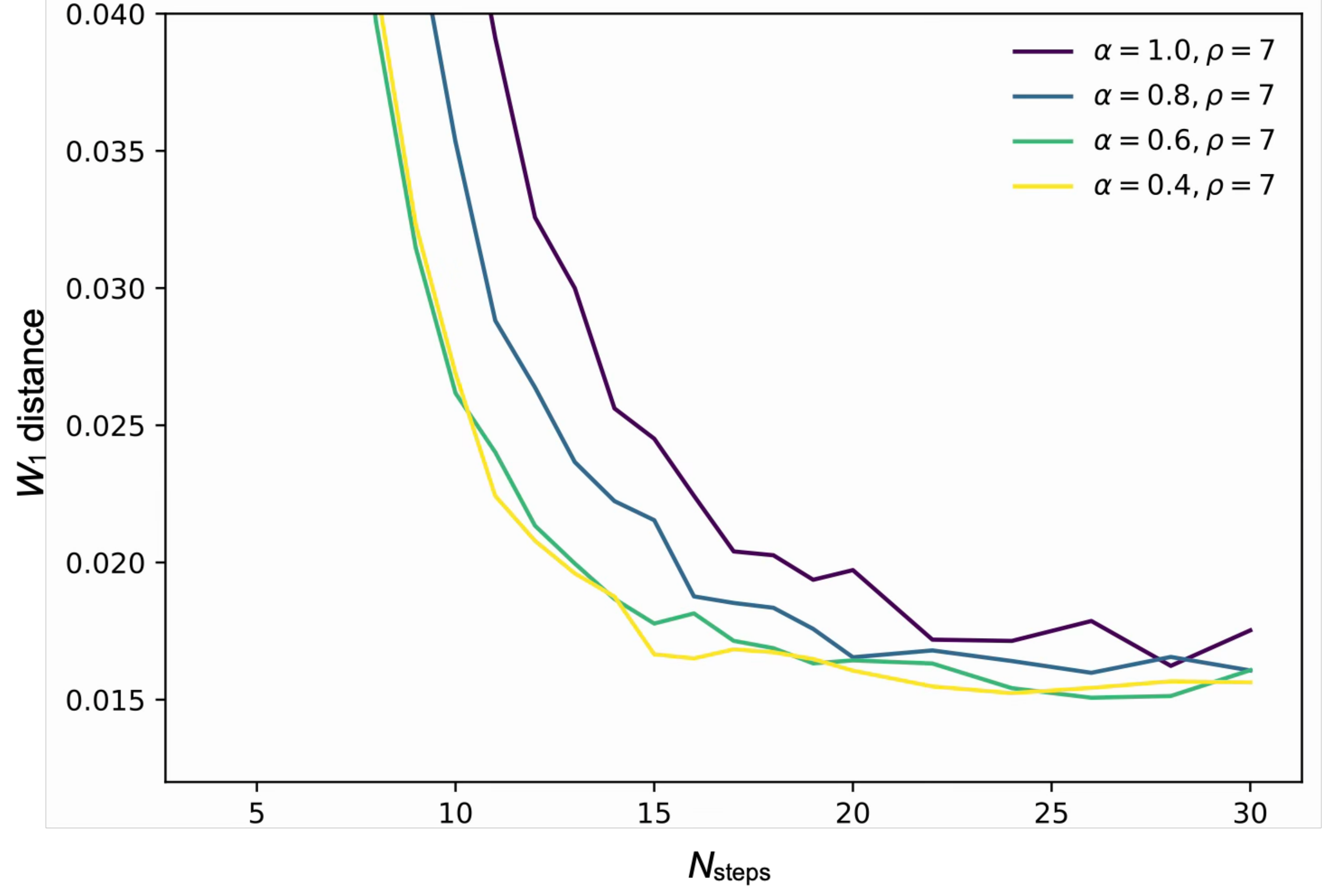}
    \caption{
    Wasserstein-1 distance, a measure of the discrepancy between probability distributions, between the generated and target galaxy distributions in the two-dimensional $M_\star$--SFR plane as a function of the number of sampling steps. Different curves show results for $\alpha = 1.0$, 0.8, 0.6, and 0.4. Smaller values indicate better agreement with the reference distribution.
    }
    \label{fig:w1_distance}
\end{figure}

We evaluated the generation quality on the validation data using several statistical measures, including the distributions of SFR and stellar mass, and the power spectrum within each patch. 
The power spectrum is computed after converting the galaxy distribution into grid data with a resolution of $(1.2~\rm Mpc)^3$. 
The galaxy power spectrum is reproduced with comparable accuracy for most choices of $\alpha$, although $\alpha = 0.4$ gives slightly worse agreement, likely because it skips too many small-$\sigma$ steps and thus fails to recover small-scale power. In contrast, the SFR and stellar mass distributions are better reproduced for $\alpha = 0.6$ and $\alpha = 0.4$, as shown in \cref{fig:w1_distance}, possibly because these global distributions are less sensitive to small-scale structure and benefit more from sampling around $\sigma \sim \sigma_{\rm data}$. We therefore adopt $\alpha = 0.6$.

\bsp	
\label{lastpage}
\end{document}